# Spatial scaling in fracture propagation in dilute systems

P. Ray and G. Date

*The Institute of Mathematical Sciences, C. I. T. Campus, Madras 600 113, India*

## Abstract

The geometry of fracture patterns in a dilute elastic network is explored using molecular dynamics simulation. The network in two dimensions is subjected to a uniform strain which drives the fracture to develop by the growth and coalescence of the vacancy clusters in the network. For strong dilution, it has been shown earlier that there exists a characteristic time $t_c$ at which a dynamical transition occurs with a power law divergence (with the exponent $z$) of the average cluster size. Close to $t_c$, the growth of the clusters is scale-invariant in time and satisfies a dynamical scaling law. This paper shows that the cluster growth near $t_c$ also exhibits spatial scaling in addition to the temporal scaling. As fracture develops with time, the connectivity length $\xi$ of the clusters increses and diverges at $t_c$ as $\xi \sim (t_c - t)^{-\nu}$, with $\nu = 0.83 \pm 0.06$. As a result of the scale-invariant growth, the vacancy clusters attain a fractal structure at $t_c$ with an effective dimensionality $d_f \sim 1.85 \pm 0.05$. These values are independent (within the limit of statistical error) of the concentration (provided it is sufficiently high) with which the network is diluted to begin with. Moreover, the values are very different from the corresponding values in qualitatively similar phenomena suggesting a different universality class of the problem. The values of $\nu$ and $d_f$ supports the scaling relation $z = \nu d_f$ with the value of $z$ obtained before.

# 1. INTRODUCTION

Disoder or inhomogeneity in the microstructure plays a very central role in the development of fracture in a material. A wide variety of disorder exists in the form of micro-defects, like grain boundaries, vacancies, inclusions, dislocations etc. which can initiate fracture in a substance. The mode of fracture can also be very different - cleavage, ductile fracture, brittle fracture, fatigue etc. The fracture process, hence, has a rich phenomenology extending over a wide range of length scales from the atomic level to the overall size of the system (see e.g. [1]). Over the past few years, lot of studies have been done to understand the role of disorder on the process of fracture in general. However, the random position of the disorder and their arbitrary shapes, dimensions and strengths pertinent to the experimental and engineering samples make the problem extremely difficult to analyze from first principles. Statistical models and concepts of statistical mechanics have turned out to be remarkably useful in this respect, and has become a powerful tool in fracture study in recent years [2].

In this letter, we study fracture in one such model which is the model of dilute elastic network. The idea behind the model is that when materials are looked on a certain length scale $L$, the development of fracture can be mimicked by a elastic network with dilution. $L$, for instance, could be the typical length separation between the elastically participating units in the material. The elastic nature of the network provides a time scale in the problem of the order of the inverse of the characteristic frequency $\sqrt{(mass/springconstant)}$. Further, the elastic bonds in the model are stretchable only upto a given limiting value after which the bonds snap irreversibly. The network is subjected to a uniform strain and disorder is introduced by removing some of the bonds randomly with probability $p$. The dilution process creates domains of vacancies in the network which, under the strain, grows and coalesce to form bigger domains by breaking further bonds and fracture develops in the system. The static properties of fracture have been studied much using this model [2]. Though such studies have been useful in answering questions like how the breaking stress varies with the dilution concentration etc., they leave the important questions, like the time taken for the fracture to propagate or the mode of propagation of fracture, completely open. These are questions concerning the dynamical aspects of the problem and in this paper we address these questions in the context of a two dimensional dilute elastic network.

As has been shown in [3] the dynamics of fracture in this model exhibits a transition at the limit of large starting bond dilution concentration $p$ (close to the percolation threshold $p_c$ [4]). The transition takes place at a characteristic time $t_c$ from a state characterized by the presence of vacancy clusters of wide distribution in sizes to a state characterized by a single important big system-spanning cluster. As a result, the average cluster size $s_c$ diverges at $t_c$ and close to $t_c$ shows the power law behaviour

$$s_c(t) \sim (t_c - t)^{-z}. \qquad (1)$$

Near $t_c$, the fracture growth process is invariant under rescaling of size and time. Consequently, the size distribution $n(s,t)$ of the vacancy clusters follows the scaling form

$$n(s,t) \sim s^{-\omega} f(s/s_c(t)) \qquad (2)$$

where $n(s,t)$ is the probability (per bond) of a vacancy cluster of mass $s$, i.e. having $s$ cut bonds, at time $t$. The exponents $\omega$ and $z$ have the values $1.05 \pm 0.05$ and $1.64 \pm 0.1$ respectively (in two dimensions).

In this paper, we study the geometrical properties of the vacancy clusters as the clusters evolve in the network with time. We find that the growth of the clusters not only satisfy the temporal scaling (2) as has been mentioned but it also exhibits spatial scaling. The clusters grow in a scale-invariant manner near $t_c$ and attain a self-similar fractal structure at $t_c$. Fractal patterns which originate from the various non-equilibrium growth processes like dielectric breakdown, diffusion-limited aggregation (DLA) or fluid-fluid displacement in porous medium have drawn great interest in the past [5–7]. Fractals are represented by fractal dimensions $d_f$ [8] defined as

$$s \sim r_s^{d_f} \qquad (3)$$

where $r_s$ is the average linear dimension of the region which contains mass $s$. $r_s$ can be any reasonable length in the problem and we take it as the average radius of gyration of the clusters of mass $s$. The value of $d_f$ obtained is $d_f = 1.85 \pm 0.05$. This is same as the corresponding value [4] for the percolation clusters but very different from the other cluster growth models (see [5,6]) which show such kinetic critical phenomena. The cluster geometry is self-similar on all length scales only at $t_c$. For $t$ away from $t_c$, it is so only upto a certain length which can be taken as the connectivity length $\xi$ defined as

$$\xi^2(t) = \frac{2\sum_s r_s^2 s^2 n(s,t)}{\sum_s s^2 n(s,t)} \qquad (4)$$

and gives the average distance of two points belonging to the same cluster. With time, larger clusters form in the network and $\xi$ increases. We find that $\xi$ diverges with a power-law $\xi \sim (t_c - t)^{-\nu}$ as $t \to t_c^-$. The exponent $\nu$ is obtained as $\nu = 0.83 \pm 0.06$.

## 2. MODEL AND SIMULATION

The molecular dynamics (MD) simulation is similar to the one used in [3]. The system is a $100 \times 100$ square elastic network which has both central as well as rotationally invariant bond-bending forces. The potential energy of the system is

$$V = \frac{a}{2} \sum_{<ij>} (\delta r_{ij})^2 g_{ij} + \frac{b}{2} \sum_{<ijk>} (\delta \theta_{ijk})^2 g_{ij} g_{jk} \qquad (5)$$

where $\delta r_{ij}$ is the change in the bond length between the nearest neighbor sites $<ij>$ from its equilibrium value $d = 1$, and $\delta\theta_{ijk}$ is the change in the bond angle between the adjacent bonds $ij$ and $jk$ from its equilibrium value $\pi/2$ which is taken to ensure the square lattice structure of the lattice at equilibrium. $g_{ij} = 1$ if the bond $ij$ is present and is 0 otherwise. $a$ and $b$ are the respective force constants of the central and bond-bending force terms. In terms of any length and time scales $\Delta$ and $\delta$ respectively, the equations of motion in dimensionless variables will involve the dimensionless parameters $\lambda_1 = a\delta^2/m$ and $\lambda_2 = b\delta^2/m\Delta^2$, where $m$ is the mass associated to the lattice sites. The parameters $\lambda_1$ and $\lambda_2$ are artefacts of our choice of scales $\Delta$ and $\delta$ and these are the parameters that enter in the simulation. It is natural to choose $\Delta$ as the lattice spacing. The ratio $\lambda_2/\lambda_1 = b/a\Delta^2$ is then a characteristic of the system under consideration. This simple dimensional analysis shows that the behaviour of the model should remain identical for the line $\lambda_2 = constant \times \lambda_1$; each particular constant modelling some particular system. One can then choose any $\lambda_1$ on such a line which may be convenient.

We have chosen $\lambda_1 = 1$ and $\lambda_2 = 0.1$ respectively (see [9] for details of the choice). The network is at first (prior to dilution) strained uniformly at all directions (the strain $l_c = 0.1$) which is then kept fixed throughout the simulation. In the next step, dilution is introduced by putting $g_{ij} = 0$ for the bonds with a probability $p$. The network is then allowed to evolve dynamically using Verlet's algorithm [10] for the updating scheme. In this scheme, the position $\vec{r}_i(t)$ of the site $i$ at time $t$ is updated to its position at later time $t + \epsilon$ according to

$$\vec{r}_i(t+\epsilon) = 2\vec{r}_i(t) - \vec{r}_i(t-\epsilon) + \vec{F}_i \epsilon^2 \qquad (6)$$

where $\vec{F}_i$ is the force on site $i$ at time $t$. The simulation involves discrete time $t$ in steps of $\epsilon$. If the simulation runs for $n$ iterations then the elapsed time $t = \epsilon n$ while the real time elapsed is $\epsilon n \delta$. One way to speed up the relaxation process (6) would be to choose a large value for $\epsilon$. However, there is an upper limit to this value for the iteration process to remain numerically stable. This limit is proportional to the convergence time for the fastest developing components of the stress distribution and this is generally very small in disordered systems. We have taken $\epsilon = 0.01$. In the course of evolution, if any bond stretches beyond a cut-off value $d_c$ (we have chosen $d_c = 1.2$), the bond snaps irreversibly and $g_{ij}$ for that bond is put to be zero. This is how fracture propagates through the network by breaking bonds. Results are obtained mostly for $p = 0.4$ (the percolation threshold for the network is $p_c = 0.5$), but few results for other values of $p$ (like $p = 0.35$ and $0.3$) are also taken to check any $p$ dependence of the results. For $p = 0.4$, the simulation is repeated for 20 different initial configurations of bond dilution to take the average over the disorder.

3. RESULTS

In the simulation, for a particular bond dilution configuration, $t_c$ is determined as the time when a vacancy cluster spanning over the network first appears. Accurate estimate of $t_c$, however, needs good disorder averaging and the precise time when the spanning cluster is first formed. In the simulation, the information regarding the clusters is recorded only at some time interval, the estimate of $t_c$ remains indeterministic to the extent of the range of the interval. The error in our estimate of $t_c$ (=7500) is ~ 10 %. Fig . 1 illustrates the network, near $t_c$, with domains of cut bonds.

The time sequence fig . 1a, b and c indicates the dynamic scaling phenomena associated with the process - the fracture pattern at later times look statistically similar to those at earlier times apart from a global increase of scale. This scale-invariance is the basis of the relation (2). On the other hand, if we compare fig . 1c and d, the former is the cluster structure at $t_c$ and the later is the blow up of the central part in c with scale factor 4, we find that the two structures again look identical in the statistical sense. This directly implies the self-similar geometry of the cluster structure.

To determine the radius of gyration $r$ of a cluster, the dual lattice of the network is considered and only those bonds are kept on the dual lattice corresponding to which there is a cut bond in the original network. It is to be noted that there is a one-to one correspondence between the bonds on the dual lattice to the bonds in the original lattice. The sites on the dual lattice which are connected by the bonds are then considered for further calculations. In this way the bond problem on the original lattice is mapped on to the more convenient site problem on the dual lattice. The number of sites $s$ in a cluster on dual lattice gives the mass $s$ and its radius of gyration is given by

$$r_s^2 = \frac{1}{s}\sum_{i=1}^{s}(\vec{r_i} - \vec{R_s})^2 \qquad (7)$$

where $\vec{R_s} = \frac{1}{s}\sum_{i=1}^{s}\vec{r_i}$ is the position vector of the centre of mass of the cluster and $\vec{r_i}$ is the position vector of the site $i$ in the cluster. Finally, $r_s$ is determined by taking the average radius of gyration over all the clusters of same mass $s$. The data are binned, where for different neighbouring cluster masses, the corresponding $r_s$-values are combined in one bin with the bin size increasing exponentially. For example, $n$-th bin contains sum of the $r_s$-values for $s$ in the range $2^{n-1}$ to $(2^n - 1)$ and the average $r_s$ is then plotted against the geometric mean of the two border sizes.

Fig. 2 shows the variation of $r_s$ with $s$ in the logarithmic scale for various time $t$ near $t_c$ in the updating process. Apart from the very small values of $s$, (3) is satisfied. The inverse of the gradient of the least square straight line fit gives $d_f = 1.85 \pm 0.05$. From the figure it is clear that as $t$ approaches $t_c$, larger and larger clusters fit into the form (3). At $t > t_c$, a very large cluster, much larger than the rest, is present in the system (see the data points for $t = 9000$) and the mass and size of this cluster deviates from the relation (3). This is the system-spanning cluster whose growth is no longer scale-invariant. This cluster is a simply connected one for which $d_f$ is close to the dimension of the space. In the simulation, $\xi$ is calculated using (4).

The variation of $\xi$ with $(t_c - t)$ is shown in Fig. 3 in the logarithmic scale. The dashed line corresponds to the least square fit of the data and its slope gives the exponent $\nu = 0.83 \pm 0.06$ which characterizes the power-law growth $\xi \sim (t_c - t)^{-\nu}$. The uncertainty in the value of $d_f$ and $\nu$ are only statistical. While we have presented results for $\lambda_2 = 0.1$, the results are similar for other values of $\lambda_2$ as well. In particular, although $t_c$ varies with $\lambda_2$, its existence is unaffected so also the spatial scaling and the fractal properties.

4. CONCLUSIONS

We explicitly show the presence of spatial scale-invariance in fracture growth in a strongly diluted system undergoing fracture. The characteristic exponents $d_f$ and $\nu$ are obtained. The value of the fractal dimension $d_f = 1.85 \pm 0.05$ is same (within the limit of statistical errors) as that for the ordinary percolation clusters but very different from the corresponding values in other cluster growth models [5,6] which exhibits such kinetic critical phenomena. The value of the connectivity length exponent $\nu \sim 0.83$. This sublinear growth of the connectivity length may be attributed to the slow stress relaxation at every length scale at this strong disorder limit. (1), (2), (3) and (4) suggest the relation $z = \nu d_f$ and the values of $\nu$ and $d_f$ satisfy this relation with the value of $z$ obtained previously. For a given value of $\lambda_2$ and in units of $\delta$, one can compute $t_c$ and this can have real applications and can be verified experimentally. It would be interesting to analyse the model for very large or very small values of $\lambda_2$ which correspond to the electrical breakdown and breakdown in central-force network respectively [9].

We would also like to note that in our modelling of fracture process a bulk relaxation time does not enter in the simulation. The dynamical simulation is the relaxation process under a constant applied strain. One could as well consider a variation with a time dependent applied stress or strain. In fact, such a variation is implicit in the studies [11] of rupture in quenched heterogeneous systems under pressure and in earthquake phenomena where an interesting log-periodic correction to scaling is observed.

## 5. ACKNOWLEDGEMENTS

FIGURES

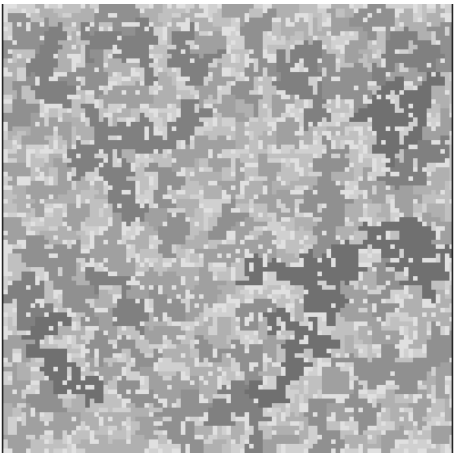 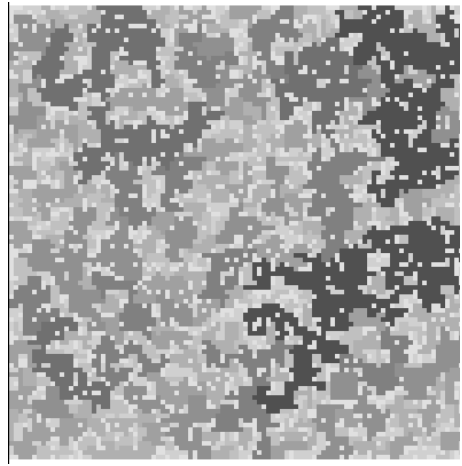

(a)                                      (b)

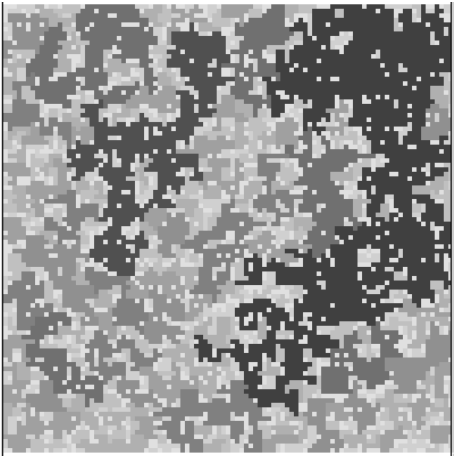 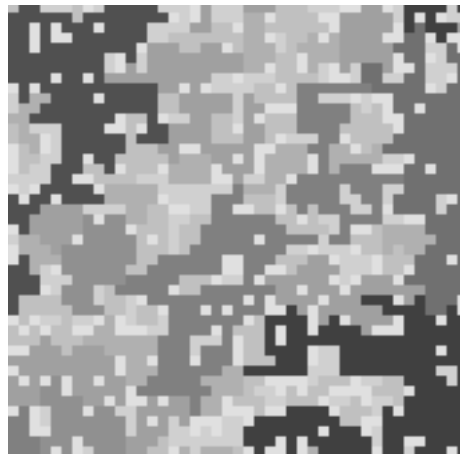

(c)                                      (d)

FIG. 1. Simulation results of fracture growth. The system size is $100 \times 100$ and the snapshots correspond to the time a) $t = 4500$, b) $t = 6000$, c) $t = 7500$. The snapshot d is the blow up of the central region in c with scale factor 4. The darker cluster represents the bigger cluster and a lighter cluster within a darker one represents a cluster within a cluster.

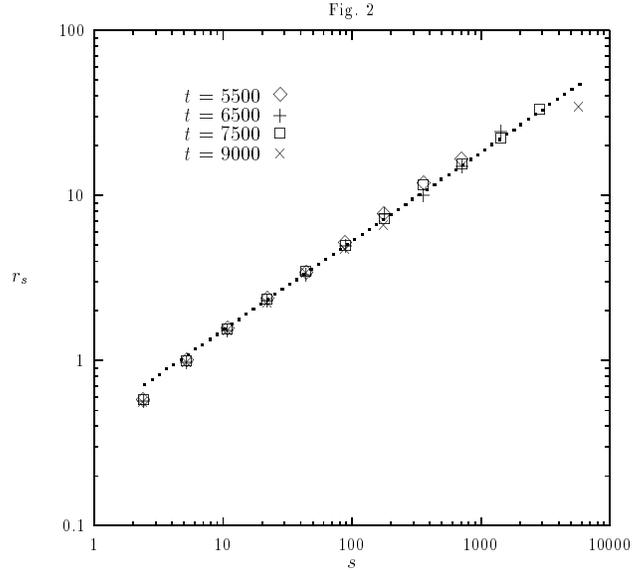

FIG. 2. The average radius of gyration $r_s$ of the clusters of mass $s$ is plotted against $s$ in the logarithmic scale. The result is for $p = 0.4$ and for various values of $t$. The dashed line has a slope $\beta = 0.54$.

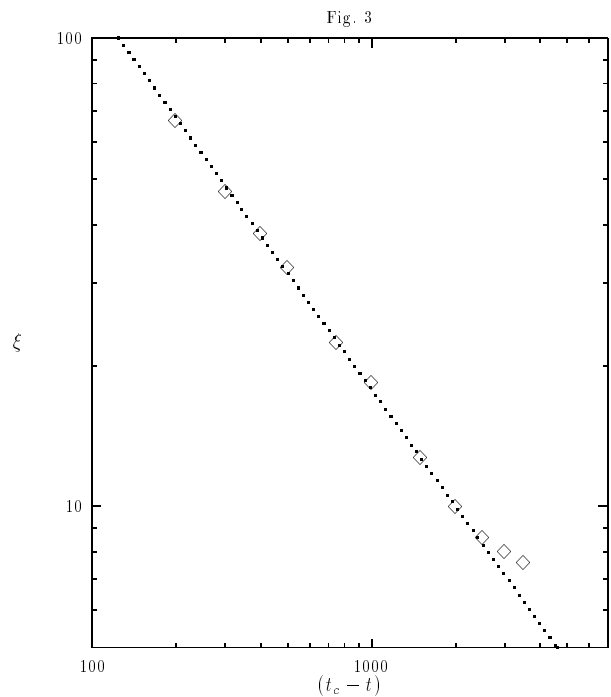

FIG. 3. The average connectivity length $\xi$ of the clusters as a function of $(t_c - t)$ for $p = 0.4$. The dashed line corresponds to the slope $\nu = 0.83 \pm 0.06$ and indicate a power-law dependence of $\xi$ on $(t_c - t)$.